# Origin of the Superior Conductivity of Perovskite Ba(Sr)SnO$_3$

Heng-Rui Liu,[1] Ji-Hui Yang,[1] H. J. Xiang,[1] X. G. Gong,[1] and Su-Huai Wei[2]

[1]Key Laboratory for Computational Physical Sciences (MOE), State Key Laboratory of Surface Physics, and Department of Physics, Fudan University, Shanghai 200433, China
[2]National Renewable Energy Laboratory, Golden, Colorado 80401, USA

**ABSTRACT**

ASnO$_3$ (A=Ba, Sr) are unique perovskite oxides in that they have superior electron conductivity despite their wide optical band gaps. Using first-principles band structure calculations, we show that the small electron effective masses, thus, good electron conductivity of ASnO$_3$ can be attributed to the large size of Sn in this system that gives the conduction band edge with antibonding Sn and O $s$ characters. Moreover, we show that ASnO$_3$ can be easily doped by La with shallow La$_A$(+/0) donor level. Our results, therefore, explained why the perovskite BaSnO$_3$, SrSnO$_3$, and their alloys are promising candidates for transparent conducting oxides.

Perovskite-structured ABO$_3$ oxides [Fig. 1(a)] have many interesting physical properties, such as high-transition-temperature superconductivity, ferroelectricity, piezoelectricity, and photoelectrochemical sensitivity; therefore, they are widely used in many technology applications.[1] Recently, there has also been great interest in applying perovskite oxides for optoelectronic applications.[2-7] For this purpose, it is desirable to have transparent conducting oxides (TCOs) based on perovskite ABO$_3$ materials.[8-18]

To be a good TCO, a material should have good optical transparency as well as high carrier mobility.[19-22] In the past, SrTiO$_3$ has been tested as an n-type TCO material because it possesses a wide optical band gap. However, its conduction band minimum (CBM) is derived mostly from the localized Ti 3$d$ states,[23] which leads to a heavy electron effective mass and consequently low carrier mobility in n-type doped SrTiO$_3$.[24] In order to design a perovskite TCO with better electron mobility, one has to modify the CBM of SrTiO$_3$ so that the CBM contains more delocalized atomic characters such as the $s$ character. One way to approach this would be substituting Ti at the B site in SrTiO$_3$ by another isovalent group IV element X that has low $s$ orbital energy and large atomic size. A good choice for this would be to replace Ti with Sn, because for Sn the 4$d$ orbitals are fully occupied and acting as semicore states, and the unoccupied 5$d$ states are very high in energy [Fig. 1(b)]. Furthermore, due to the large atomic size of Sn, the CBM, which is an antibonding Sn and O $s$ state, will be low in energy. Consequently, following the doping limit rule, this will make the compounds more easily doped n-type.[25,26]

To test these ideas, in this letter, we investigate the structural, electronic, and doping properties of ASnO$_3$ with A = Ba, Sr. For comparison, we also compare the results with A site occupied by group II-B elements Cd and Zn and B site occupied by transition metals such as Ti. Our results show that both BaSnO$_3$ and SrSnO$_3$ have wide optical gaps and small electron effective masses. Moreover, the (+/0) transition energy levels of the La$_A$-doped BaSnO$_3$ and SrSnO$_3$ are both very shallow, so the materials can be easily doped n-type. Our results, thus, confirmed that BaSnO$_3$, SrSnO$_3$, and their alloys are promising TCO materials, as expected from our designing principles.

The structural and electronic properties of the cubic perovskites were calculated within the density functional formalism as implemented in the VASP[27] code. For the exchange-correlation potential, we used the standard Heyd-Scuseria-Ernzerhof (HSE06)[28] hybrid functional. The projector-augmented-wave[29] pseudopotentials with an energy cutoff of 500 eV for plane-wave basis sets were employed to give converged results. The phonon spectra calculation used the Quantum Espresso[30] code by means of plane-wave basis sets and the Vanderbilt ultra-soft pseudopotential.[31] The exchange correlation energy was evaluated by the GGA (PW91) functional,[32] and the dynamical matrices were calculated on a uniform 5x5x5 grid of q-points. The vibrational properties were computed using density functional perturbation theory (DFPT).[33]

The calculated phonon spectrum [Fig. 2(a)] of the BaSnO$_3$ shows that distortions around the SnO$_6$ octahedral may occur at low temperature, which is consistent with experimental findings.[18,34] The situation of the SrSnO$_3$ is similar to that of the BaSnO$_3$. By distorting the cubic phase following the soft phonon mode at M and one of the equivalent threefold degenerate modes at R, we get two distorted phases, that is, the distorted_M phase with the space group P4/mbm, and the distorted_R phase with the space group R-3c, respectively [Fig. 2 (b)]. For the distorted phases, the deviations from the cubic phase are small, and the total energies of the three phases are close to each other: the distorted M phase and the distorted R phase are about 2 (25) meV/atom and 7 (34) meV/atom more stable than the cubic Ba(Sr)SnO$_3$ phase, respectively. The larger stability and distortion of the distorted SrSnO$_3$ is consistent with the observation that when the atomic size of A decreases, the distortion increases.[21]

The band structures of the cubic perovskite BaSnO$_3$ and SrSnO$_3$ are similar [Figs. 3(a) and 3(b)]. From group theory, we find that the electric-dipole transition at Γ from the VBM to the CBM is allowed, which leads to an optical band gap of 3.0 (4.04) eV for Ba(Sr)SnO$_3$. Moreover, the electric-dipole transition at Γ from the CBM to the second conduction band in both materials is forbidden, which is important for the transparency when the material is heavily n-type doped.[19] The distorted structures both have 10 atoms per primitive cell and their band structures [Figs. 3(c) and 3(d)] can be viewed as resulting from the folding of that of the cubic phase plus small distortions. For example, the M point of the cubic phase is folded to the Γ point for the distorted_M phase, which makes the distorted M phase a direct gap material with a fundamental gap of 2.59 eV. However, the optical properties are almost the same in these structures. Therefore, in the following, we will focus only on the cubic phase.

The total and partial density of states (DOS) of the cubic perovskite BaSnO$_3$ are plotted in Fig. 4(a), the situation of SrSnO$_3$ is qualitatively the same and not shown. It is clear that the upper valence bands are mainly derived from the $p$ states of O atoms, while the lower conduction bands are mainly derived from the hybridization between the $s$ and $p$ states from Sn atoms and O atoms. The VBM is completely comprised of the O $p$ characters and is thus a non-bonding state, while the CBM is the antibonding Γ$_1$ state derived from the hybridization between the $s$ states of Sn atoms and O atoms [Fig. 4(b)].

By substituting the cations at A and B sites of SrTiO$_3$, one can obtain other cubic perovskite oxides. To understand the substitution effects, we calculate the optical gaps and electron effective masses of various cubic perovskites, and the results are summarized in Table I. The CBM of these materials are all at the Γ point, and the electron effective mass tensor is isotropic with respect to the cubic perovskite x, y, and z axes guaranteed by the O$_h$ point group symmetry. The studied perovskites here can be divided into three groups according to the composition of cations: the first and the second group both have an Sn atom at the B site, while the A site is occupied by group II-A alkaline earth metals in the first group and by group II-B post transition metals (TMs) in the second group, respectively. We also list the results for SrTiO$_3$ and another common TM perovskite, LaAlO$_3$, in the third group for comparison.

The difference in cations leads to different character of the CBM in these perovskites: the CBM is the antibonding $s$ state from the Sn-O hybridization and the antibonding *s* state from the A-O (A = Zn, Cd) hybridization for perovskite in the first and the second group, respectively, while *d* states dominate the CBM for the TM perovskites. For perovskites in the first group, the strength of the Sn-O hybridization increases from $BaSnO_3$ to $CaSnO_3$ due to the shorter Sn-O bond length caused by the decrease of the size of the A site cations. Because the CBM is an anti-bonding state of the hybridization, it will be pushed up in energy, leading to more flat band dispersion when the coupling enhances, which explains the increase of the band gap and the electron effective mass from the $BaSnO_3$ to $CaSnO_3$. The A site cations of the perovskites in the second group have much lower *s* state energies compared to those in the first group, and the Sn-O hybridization is stronger due to the smaller cation radius. As a result, the anti-bonding *s* state from the Sn-O hybridization is above the antibonding A-O hybridized *s* states, so the CBM is thus dominated by the A-O *s* states. Furthermore, the stronger *p-d* repulsion in the second group compounds also raises the VBM and leads to smaller band gaps and poor optical transition compared with those in the first group.

Our results clearly show that the optical gap and the character of the CBM of the $ABO_3$ compound could be adjusted by substituting the cation at the A or B site, and small electron effective mass could be obtained when the CBM has the *s* character resulted from either the A-O or the B-O hybridization. On the other hand, the localized *d* states lead to significantly heavier effective mass in TM perovskites. $BaSnO_3$ has a very small electron effective mass of 0.2 $m_e$ and a relatively large optical band gap around 3.0 eV and is thus one of the best perovskite-structured TCOs. Its electronic and optical properties can be further tuned by alloying with other compounds such as $SrSnO_3$, which has a larger optical gap and an only slightly heavier electron effective mass compared to the $BaSnO_3$.

To be a good n-type TCO, the material should also be easily doped and the donor level should be shallow. We expect that the best n-type dopant for $BaSnO_3$ should be La substitution on the Ba site. This is because La atom has a similar size to the Ba atom and the valance difference between them is only one, so the solubility of La in $BaSnO_3$ should be high. More importantly, the La d orbital energy is high and the CBM of $BaSnO_3$ has very little contribution from Ba, so substitution on the Ba causes minimum perturbation at the CBM, consequently the $La_{Ba}(+/0)$ donor level should be very shallow. Using the supercell approach,[35] we calculate the $La_A(+/0)$ transition energy levels in a 3x3x3 cubic supercell that contains 135 atoms (with one A atom substituted by one La atom) for the cubic perovskite $ASnO_3$ (A = Ba, Sr, Ca), the results are summarized in Table II. The donor levels are similar for $BaSnO_3$ and $SrSnO_3$ and become deeper for $CaSnO_3$ when the CBM increases. The donor levels are found to be quite shallow in $BaSnO_3$ and $SrSnO_3$, which means electrons from La atoms can be easily ionized to conduction bands, making La a good n-type dopant, as expected.

In conclusion, the structural, electronic and doping properties of perovskite $BaSnO_3$ and $SrSnO_3$ are studied using first-principles methods. We find that (i) small rotations of the $SnO_6$ octahedral may occur in these compounds at low temperature, especially for $SrSnO_3$; (ii) these two perovskite oxides both have a wide optical gap, which makes them transparent in the visible region; (iii) despite the large optical gap, the electron effective masses of these two compounds are quite small due to the antibonding *s* character at their CBM states; (iv) $BaSnO_3$ and $SrSnO_3$ can be easily n-type doped, and La is an ideal n-type dopant in these materials because of the shallow donor level. These results are explained by the size and atomic wavefunction characters of the elements and suggesting that the perovskite $BaSnO_3$, $SrSnO_3$ and their alloys $(Ba,Sr)SnO_3$ are good TCO materials.

This work is partially supported by the Special Funds for Major State Basic Research, National Science Foundation of China, Ministry of Education and Shanghai Municipality. The calculations were performed in the Supercomputer Center of Fudan University. The work at NREL is supported by the U.S. Department of Energy under Contract No. DE-AC36-08GO28308.

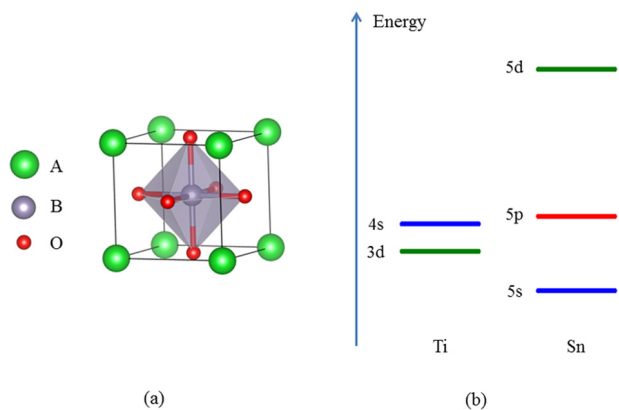

FIG. 1. (Color online) (a) Primitive cell of cubic perovskite $ABO_3$, and (b) schematic diagram of different energy levels of valence electrons in the element at the B site of cubic perovskite $ABO_3$. The Sn $4d$ orbitals (not shown) are fully occupied and are very low in energy.

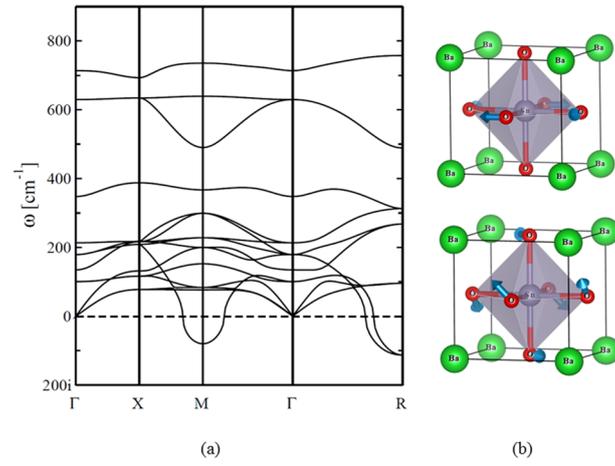

FIG. 2. (Color online) (a) The phonon dispersion relations of the cubic perovskite $BaSnO_3$, and (b) the eigenvectors (marked by blue arrows) of the soft phonon mode at M (the upper panel) and one of the equivalent threefold degenerate imaginary modes at R (the lower panel).

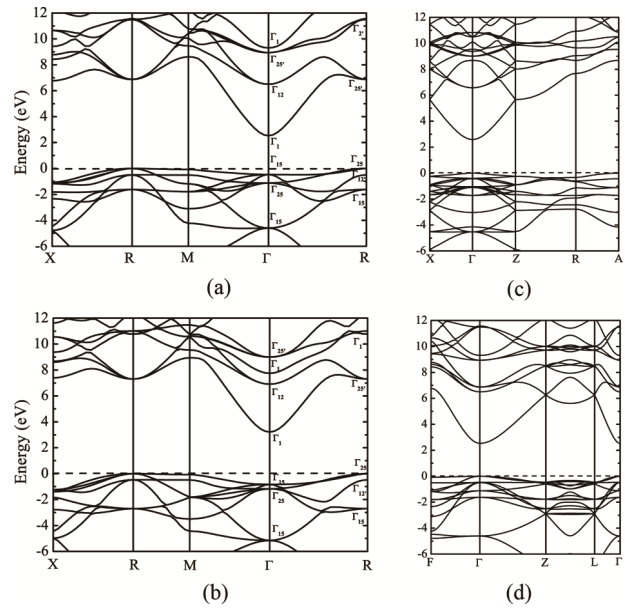

FIG. 3. Band structure for: (a) the cubic BaSnO$_3$, (b) the cubic SrSnO$_3$, (c) the distorted_M BaSnO$_3$, and (d) the distorted_R BaSnO$_3$ by the HSE06 functional. The symmetries of the bands at the $\Gamma$ and the R point of cubic perovskite Ba(Sr)SnO$_3$ are also labeled in (a) and (b).

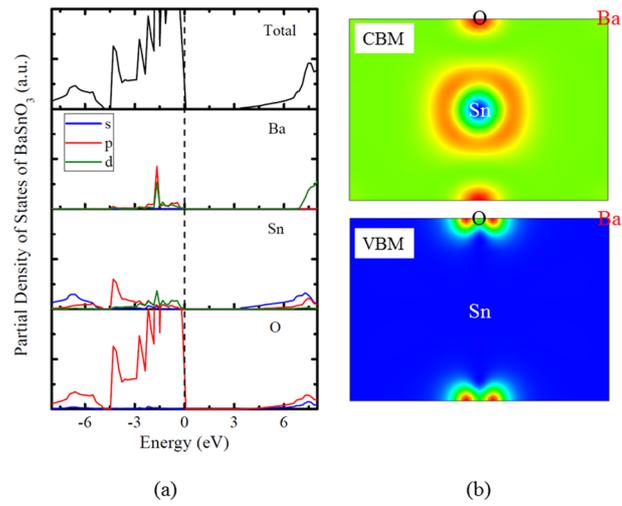

FIG. 4. (Color online) (a) Total and partial density of states (DOS), and (b) partial charge at the CBM and the VBM in the (110) plane of cubic perovskite BaSnO$_3$. The VBM is set to zero in (a).

Table I. Direct band gap $E_g$, electron effective mass m*, and the composition of the CBM at the Γ point of various cubic perovskites by the HSE06 functional. The band dispersions of SrTiO$_3$ and LaAlO$_3$ are averaged over degenerate bands around Γ.

|  | BaSnO$_3$ | SrSnO$_3$ | CaSnO$_3$ |
|---|---|---|---|
| $E_g$ (eV) | 3.0 | 4.04 | 4.62 |
| m* (m$_e$) | 0.20 | 0.23 | 0.25 |
| Composition of the CBM | s states from Sn-O hybridization | | |
|  | ZnSnO$_3$ | | CdSnO$_3$ |
| $E_g$ (eV) | 1.31 | | 2.36 |
| m* (m$_e$) | 0.17 | | 0.21 |
| Composition of the CBM | s states from A-O hybridization | | |
|  | SrTiO$_3$ | | LaAlO$_3$ |
| $E_g$ (eV) | 3.38 | | 5.04 |
| m* (m$_e$) | 0.57 | | 0.45 |
| Composition of the CBM | d states from TM-O hybridization | | |

Table 2. La$_A$(+/0) transition energy level with respect to the CBM in the II-A cubic perovskite ASnO$_3$ (A = Ba, Sr, Ca) calculated using the PBE functional.

|  | BaSnO$_3$ | SrSnO$_3$ | CaSnO$_3$ |
|---|---|---|---|
| $E_{CBM}$ - E[La$_A$(+/0)] (eV) | 0.046 | 0.033 | 0.089 |